\documentstyle[12pt,aps]{revtex}
\draft

\begin{document}

\title{{\small To appear in Journal of Mathematical Physics (2002)} \\
Extensive form of equilibrium nonextensive statistics}

\author{Qiuping A. Wang and Alain Le M\'ehaut\'e}
\address{Institut Sup\'erieur des Mat\'eriaux du Mans,\\ 44, Avenue F.A.
Bartholdi, 72000 Le Mans, France}


\maketitle

\begin{abstract}
It is argued that, in nonextensive statistical mechanics with Tsallis entropy,
the factorization of compound probability over subsystems is a consequence of
the existence of thermodynamic equilibrium in the composite system and should
be respected by all exact calculations concerning equilibrium subsystem. Using
nonadditive energy satisfying this factorization, we propose an additive
formalism of nonextensive statistical mechanics with additive $q$-deformed
physical quantities and exponential distributions. This formalism leads to
exact quantum gas distributions different from those given by factorization
approximation with additive energy. The fermion distribution of present work
shows similar characteristics to the distribution of strongly correlated
electrons given by numerical analysis with Kondo t-J model.
\end{abstract}



\section{Introduction}

In this paper we will discuss some problems of nonextensive statistical
mechanics (NSM) relevant to the factorization of compound probability into
product of single body probability :
\begin{eqnarray}                                        \label{m1}
\rho=\prod_{n=1}^N\rho_n,
\end{eqnarray}
where $N$ is the number of subsystems in the system of interest, $\rho$ is the
$q$-exponential distribution (QED) : $\rho\propto[1-(1-q)\beta
H]^{\frac{1}{1-q}}$ ($[.]\geq 0$), as given by the maximization of Tsallis
entropy $S_q=-\texttt{Tr}\frac{\rho-\rho^q}{1-q}$ (Boltzmann constant $k_B=1$
and $q>0$) under some constraints\cite{Tsal88,Marti,Wang00}. This factorization
Eq.(\ref{m1}) has been viewed as a result of the independence of noninteracting
subsystems having additive energy, just as in Boltzmann-Gibbs statistics (BGS)
supposing short range interactions, and caused confusions in some theoretical
studies of NSM and its applications to many-body problems. On the basis of a
new idea relating Eq.(\ref{m1}) to thermodynamic equilibrium, we will argue
that these confusions can be avoided if we introduce suitable nonadditive
thermodynamic variables satisfying Eq.(\ref{m1}). Some theoretical consequences
for quantum distributions of this ``equilibrium version" of NSM will be
studied.

Due to the necessity of defining additive average value of some extensive
$q$-deformed thermodynamic variables, the discussions will be made within the
formalism of incomplete statistics (IS) with $\texttt{Tr}\rho^q=1$ and
normalized average $\bar{x}=\texttt{Tr}\rho^q x$\cite{Wang00}. The reader will
find that the quantum distributions of IS indeed show some particular
properties already noticed with strongly correlated electrons.

\section{About incomplete normalization}

IS as a alternative version of NSM was originally motivated by some theoretical
peculiarities in the last Tsallis version of NSM based on the conventional
normalization and unnormalized expectations\cite{Wang00}.

The basic assumption of IS is that our knowledge about physical systems is in general incomplete due to unknown
space-time correlations or the effects of known interactions which can not be exactly studied (such as chaos).
In this case, probability distributions are incomplete, i.e., $\texttt{Tr}\rho=Q\neq 1$\cite{Reny66} (or
$\sum_{i=1}^{w}p_i=Q$ where $w$ is only the number of accessible states in phase space). One can only write
$\texttt{Tr}F(\rho)=1$ where $F$ is certain function of $\rho$. In the case of complete or approximately
complete distribution (such as in BGS), $F$ is identity function. Recently, in order to overcome some of
theoretical difficulties of NSM in keeping the framework prescribed by Tsallis entropy, we
proposed\cite{Wang00} $F(\rho)=\rho^q$ so that
\begin{equation}                                \label{3}
\texttt{Tr}\rho^q=1,
\end{equation}
where $q$ is the {\it incompleteness index}\cite{Wang00}. Since $\rho<1$, we have to set $q\in[0,\infty]$.
$q=0$ should be avoided because it leads to $\rho=0$ for all states. We note that Eq.(\ref{3}) has been
successfully employed to deduce some power laws based on R\'enyi's entropy\cite{Bash00}.

This kind of {\it incomplete normalization} is possible whenever the phase space is partially known or
accessible. With a fractal or chaotic phase space, e.g., a complete calculation of probability becomes in
general impossible. In this sense, a plausible justification of Eq.(\ref{3}) may be inspired by a work of
Tsallis\cite{Tsal95} discussing nonadditive energy and probability distributions on fractal supports, although
at that stage the work was not connected to anomalous normalization like Eq.(\ref{3}). In that work,
considering some simple self-similar fractal structures (e.g. Cantor set), one can obtain :
\begin{equation}                                \label{3a}
\sum_{i=1}^{W}[\frac{V_i(k)}{V(0)}]^{d_f/d}=1
\end{equation}
where $V_i(k)$ may be seen as the segments of the fractal structure at a given
iteration of order $k$, $V(0)$ a characteristic volume of the fractal structure
embedded in a $d$-dimension Euclidean space, $d_f=\frac{\ln n}{\ln m}$ is the
fractal dimension, $n$ the number of segments replacing a segment of the
precedent iteration, $m$ the scale factor of the iterations and $W=n^k$ the
total number of segments at the $k^{th}$ iteration. If we suppose that the
fractal structure with $k\rightarrow \infty$ is a phase space containing
homogeneously distributed points, the {\it exact microcanonical probability
distribution} of the $k^{th}$ iteration can be defined as
$$p_i=\frac{V_i(k)}{V}=\frac{V_i(k)}{\sum_i^WV_i(k)}$$ where $V$ is the total
volume of the phase space. This distribution obviously sums to one. The problem is that $V$ is an indefinite
volume as $k\rightarrow \infty$ and impertinent for exact probability definition. In addition, $V$ is not
differentiable and contains inaccessible points. Thus exact summation in $V$ would be impossible. Now if we
define $p_i=\frac{V_i(k)}{V_0}$ as a physical or effective distribution, then we have
$$\sum_{i=1}^{W}p_i^{d_f/d}[V_0/V(0)]^{d_f/d}=1,$$ where $V_0$ is {\it a completely accessible and infinitely
differentiable support} on which the calculation of $p_i$ is possible. If we choose $V_0=V(0)$, we can write
Eq.(\ref{3}) with $q=d_f/d$. The conventional normalization $\sum_{i=1}^{W}p_i=1$ can be recovered when
$d_f=d$.

The above example is only a case of equiprobable distribution on simple fractal
structure, but it illustrates very well the possibility that, in complex cases,
a physical probability may not sum to one and may sum to unity only through a
kind of power normalization, which, pertinent and useful for incomplete
distributions, is consistent with the discussions of reference \cite{Tsal95} on
the mass calculation and the information consideration in porous structures.

\section{Factorization of compound probability and thermodynamic equilibrium}

In NSM, there are two major problems connected tightly to the factorization of compound probability. The first
concerns the application of NSM to many-body systems via one-body distribution. NSM is originally intended to
describe complex systems with long range interactions or fractal structure of space-time showing nonextensive
phenomena. So from the beginning of this theory, Eq.(\ref{m1}) is supposed for composite systems containing $N$
{\it statistically independent subsystems} in order to elicit the nonextensive character by following relation
:
\begin{eqnarray}                                        \label{m2}
\ln[1+(1-q) S_q]=\sum_{n=1}^N \ln[1+(1-q) S_q(n)]
\end{eqnarray}
For $N=2$, $S_q=S_q(1)+S_q(2)+(1-q) S_q(1)S_q(2)$ as one often finds in the
literature. Due to this $independence$, it has been believed by many that exact
calculations within NSM should use the additive hamiltonian $H_0=\sum_{n=1}^N
H_n$, where $H_n$ is the hamiltonian of $n^{th}$
subsystem\cite{Marti,Tsal94,Prat94,Curi96,Lenz98,Lenz01}. However, this
hamiltonian is not compatible with neither Eq.(\ref{m1}) nor Eq.(\ref{m2})
since these equations applied to QED mean\cite{Tsal88,Wang01,Beck00}:
\begin{eqnarray}                                        \label{2}
H &=& \sum_{n=1}^N H_n + \sum_{k=2}^N [(q-1)\beta]^{k-1}
    \sum_{n_1<n_2<...<n_k}^N\prod_{j=1}^k H_{n_j} \\ \nonumber
&=& H_0+H_c,
\end{eqnarray}
where $\beta$ is the inverse temperature. In order to reconcile $H_0$ and
Eq.(\ref{m1}), a so called {\it factorization approximation} is
proposed\cite{Buyu93} by neglecting the second term at the right hand side of
Eq.(\ref{2}). This approximation has been, explicitly or not, employed in most
of the applications of NSM\cite{Beck00,Buyu93,Kani96,Bedi01,Beck01a,Tsal01} via
one-body QED. These applications certainly shows the usefulness of one-body
QED, but the approximation neglecting the correlation energy by supposing
sometimes weak interacting dilute particles\cite{Beck01} is not a reassuring
basis. Indeed, some recent works show that the correlation energy ($H_c$) given
by the second term of Eq.(\ref{2}) is in general not negligible\cite{Wang01}
and that the partition function given by using additive energy is completely
different from that given by using Eq.(\ref{2}) when $N$ is large\cite{Lenz01}.
So a doubt arises about the connection between independence of subsystems and
additive energy. Recently, an interesting idea is forwarded to define a
 ``quasi-independence" according to nonadditive energy Eq.(\ref{2}) in order to
apply NSM to turbulence flow problems\cite{Beck01}. As a matter of fact, this
proposal implies rejection of classical independence for Eq.(\ref{m1}).

The second problem connected to probability factorization is the establishment
of zeroth law and the definition of temperature for NSM. It was believed that
the zeroth law of thermodynamics was absent within NSM\cite{Guer96} due to the
paradox between Eq.(\ref{m1}) and the additive energy. Recently, a series of
works have been published on this issue\cite{Abe00} claiming the establishment
of zeroth law and the definition of a generalized temperature on the basis of
additive hamiltonian $H_0$ and Eq.(\ref{m2}) by neglecting $H_c$. It is evident
that the paradox mentioned above persists behind this approximate zeroth law.

The central question is : Eq.(\ref{m1}) certainly implies independence of
noninteracting systems for $BGS$, but does it mean the same thing for NSM? Very
recently, Abe\cite{Abe01} proposed a general pseudoadditivity for entropy
required by the existence of thermal equilibrium in composite nonextensive
systems. For a system containing $N$ subsystems, the pseudoadditivity is :
\begin{eqnarray}                                        \label{2b}
\ln[1+\lambda_S f(S)]=\sum_{n=1}^N\ln[1+\lambda_S f(S_n)],
\end{eqnarray}
where $f$ is certain differentiable function satisfying $f(0)=0$ and
$\lambda_S$ a constant depending on the nature of the system of interest. On
the other hand, Eq.(\ref{2b}) applied to Tsallis entropy means $f(S)=S$ and
$\lambda_S=1-q$\cite{Abe01}, which directly leads to
$\ln\texttt{Tr}\rho^q=\sum_{i=1}^N\ln\texttt{Tr}\rho_i^q$ or Eq.(\ref{m1})
(i.e. with classical probability $p_i$ of the state $i$, $(p_ip_j)^q=p_{ij}^q$
means $p_ip_j=p_{ij}$). So Eq.(\ref{m1}) {\it has nothing to do with
statistical independence of subsystems}. It is {\it a consequence of the
existence of thermodynamic equilibrium} and must be rigorously respected by all
exact calculations. Equilibrium energy has been proved\cite{Wang02} to satisfy
the same kind of pseudoadditivity as Eq.(\ref{2b}) ($S$ is replaced by $H$). If
we choose $f(H)=H$ and $\lambda_H=(q-1)\beta$, we get
\begin{eqnarray}                                        \label{2c}
\ln[1+(q-1)\beta H]=\sum_{n=1}^N\ln[1+(q-1)\beta H_n]
\end{eqnarray}
which is just Eq.(\ref{2}) satisfying Eq.(\ref{m1}). In this way, the zeroth
law becomes evident and a temperature can be straightforwardly defined at
maximum entropy and minimum energy\cite{Wang00,Wang01}.

\section{Additive formalism of NSM}

\subsection{Information measure}
The $g$-logarithmic information measure
\begin{eqnarray}                                        \label{4}
I_\nu=\frac{(1/\rho)^\nu-1}{\nu}
\end{eqnarray}
is a nonadditive generalization of Hartley formula $I=\ln(1/\rho)$ and can be
employed to deduce Tsallis entropy \cite{Tsal88,Wang00,Tsal01a}. $I_g$ or $I$
is the information needed to specify at which state the system is localized.
$\nu$ equals $1-q$ or $q-1$, depending on the normalization procedures of
$\rho$\cite{Entropy}. Using Eq.(\ref{m1}), we get :
\begin{eqnarray}                                        \label{6}
\ln(1+\nu I_\nu)=\sum_{n=1}^N \ln(1+\nu I_\nu^{(n)})
\end{eqnarray}
where $I_\nu^{(n)}$ is the information needed to specify the $n^{th}$
subsystem. This pseudoadditivity is evident if we recast the generalized
Hartley formula Eq.(\ref{4}) as follows
\begin{eqnarray}                                        \label{7}
I_\nu=\frac{e^{-\nu\ln\rho}-1}{\nu}=\frac{e^{\nu I}-1}{\nu}
\end{eqnarray}
where
\begin{eqnarray}                                        \label{8}
I=\ln\frac{1}{\rho}=\frac{\ln(1+\nu I_\nu)}{\nu}.
\end{eqnarray}
can be referred to as $q$-deformed information measure and is additive supposed
Eq.(\ref{m1}). It is noteworthy that this $I$ is not the quantity of Hartley
information if $\rho$ is a nonextensive distribution for $\nu\neq 0$.

\subsection{Canonical ensemble}
Now let us define an additive entropy $S$ as follows :
\begin{eqnarray}                                        \label{9}
S=\texttt{Tr}\rho^q\ln\frac{1}{\rho}
\end{eqnarray}
and an additive $q$-deformed ``hamiltonian"
\begin{eqnarray}                                        \label{11b}
h=\frac{\ln[1+(q-1)\beta H]}{(q-1)\beta}.
\end{eqnarray}
So Eq.(\ref{2c}) becomes
\begin{eqnarray}                                        \label{12}
h=\sum_{n=1}^Nh_n.
\end{eqnarray}
This means following transformations :
\begin{eqnarray}                                        \label{13}
H=\frac{e^{(q-1)\beta h}-1}{(q-1)\beta }, \;\;\; \;\;\;H_n=\frac{e^{(q-1)\beta
h_n}-1}{(q-1)\beta }
\end{eqnarray}
and
\begin{eqnarray}                                        \label{14}
\rho=\frac{1}{Z}[1+(q-1)\beta H]^{1/(1-q)}=\frac{1}{Z}e^{-\beta h},
\end{eqnarray}
where $Z^q=\texttt{Tr}e^{-q\beta h}$\cite{Wang00}. It should be noticed that,
when addressing a system of $N$ particles, we have to write
$H_n=\frac{p_n^2}{2m}+V_n$ for single particle so that
$h_n=\frac{\ln[1+(q-1)\beta(\frac{p_n^2}{2m}+V_n)]}{(q-1)\beta}$ where
$\frac{p_n^2}{2m}$ is the classical kinetic energy and $V_n$ is the potential
energy. It is clear that $H_n$, instead of $h_n$, is the physical energy. When
$q=1$ ($H_c=0$), we recover $H_n=h_n$ and $H=h=\sum_{n=1}^N\frac{p_n^2}{2m}$.
The $q$-deformed internal energy $u$ is defined as follows :
\begin{eqnarray}                                        \label{15}
u=\texttt{Tr}\rho^q h.
\end{eqnarray}
We can easily show that the distribution Eq.(\ref{14}) can be yielded by the
maximum of the additive ``entropy" $S$ (which surely exists due to the
monotonic relation between $I$ and $I_\nu$) under the constraints of
Eq.(\ref{15}) and incomplete normalization $\texttt{Tr}\rho^q=1$. It is easy to
verify that $S=\ln Z+\beta u$ and, via the zeroth law, $\frac{\partial
S}{\partial u}=\beta=1/T$. The ``first law" is given by $du=TdS-pdV$ where $p$
is $q$-deformed pressure and $V$ the volume of the system which is chosen to be
additive here. The $q$-deformed Helmholtz free energy $f$ is defined as
$f=u-TS=-T\ln Z$ and can be connected to the nonadditive one
$F_q=-T\frac{Z^{1-q}-1}{1-q}$\cite{Tsal88,Wang00} as follows :
\begin{eqnarray}                                        \label{16}
f=\frac{\ln[1+(q-1)\beta F_q]}{(q-1)\beta}.
\end{eqnarray}
So $p=-(\frac{\partial f}{\partial V})_T=P/Z^{1-q}$ where $P=-(\frac{\partial
F_q}{\partial V})_T$ is the real pressure. In this scenario, the thermodynamic
equilibrium of a system $C$ containing two equilibrium systems $A$ and $B$
satisfying $V(C)=V(A)+V(B)$ corresponds to $\beta(A)=\beta(B)$ and $p(A)=p(B)$.
This implies that $P(A)\neq P(B)$ if $Z(A)\neq Z(B)$. This is because we have
supposed nonadditive energy and additive volume. As a matter of fact, in this
formalism, if we want that $P(A)=P(B)$ at equilibrium, we must accept
nonadditive volume and additive $q$-deformed volume $v$ with which the first
law becomes $du=TdS-Pdv$. This means : $P=-(\frac{\partial F_q}{\partial
V})_T=-(\frac{\partial f}{\partial v})_T$. This relation can help to deduce
$v-V$ relation. We also have $v=(\frac{\partial g}{\partial P})_T$ where the
$q$-deformed Gibbs energy is given by $g=f+Pv$. Since $v$ is additive, $V$ will
be nonadditive if $v$ is not a linear function of $V$. We will come back to
this issue later in this paper.

\subsection{Grand canonical ensemble}
It is known that the grand canonical ensemble $QED$ has been given
by\cite{Tsal88,Curi96,Buyu93} :
\begin{eqnarray}                                        \label{17}
\rho\propto [1-(1-q)\beta(H-\mu N)]^\frac{1}{1-q}
\end{eqnarray}
for $N$ identical particle systems, where $\mu$ is chemical potential. This
distribution has been widely used for quantum particle
systems\cite{Buyu93,Kani96,Bedi01,Wang02b}. But the zeroth law has never been
rigorously established for this ensemble. In a previous work\cite{Wang02b}, one
of the authors of present paper deduced exact quantum distributions on the
basis of Eq.(\ref{17}) and following relation suggested by Eq.(\ref{m1}) :
\begin{eqnarray}                                        \label{17a}
\rho\propto [1-(1-q)\beta(H_n-\mu)]^\frac{N}{1-q}.
\end{eqnarray}
In the framework of IS\cite{Wang00,Wang02b}, the exact distributions are given
by
\begin{eqnarray}                                        \label{17b}
\bar{n}_k=\frac{1}{[1+(q-1)\beta (e_k-\mu)]^{\frac{q}{q-1}}\pm 1}
\end{eqnarray}
where $e_k$ is the energy of one-particle state $k$  and``+" and ``-"
correspond to fermions and bosons, respectively. Now we show that this
distribution can be written in exponential form just as for conventional
noninteracting quantum gases and that the zeroth law can be rigorously
verified.

Let us suppose
$\beta=\frac{\beta'}{1+(q-1)\beta'\mu}=\beta'[1-(q-1)\beta'\mu']$ (or
$\beta'=\frac{\beta}{1-(q-1)\beta\mu}$) and
$\mu=\frac{\mu'}{1+(1-q)\beta'\mu'}$ (or
$\mu'=\frac{\mu}{1+(q-1)\beta'\mu}=\mu[1-(q-1)\beta\mu]$) which imply
$\beta'\mu'=\beta\mu$. Eq.(\ref{17a}) can be recast as :
\begin{eqnarray}                                        \label{18}
\rho &=&\frac{1}{Z}[1-(1-q)\beta'H_n]^\frac{N}{1-q}[1+(1-q)\beta'\mu']^\frac{N}{1-q}
\\ \nonumber &=&\frac{1}{Z}e^{-N\beta(h_n-\omega)}
\end{eqnarray}
where $\omega=\frac{\ln[1+(1-q)\beta'\mu']}{(1-q)\beta'}$ and
$h_n=\frac{\ln[1+(q-1)\beta' H_n]}{(q-1)\beta'}$. $Z^q=\texttt{Tr}[1-(1-q)\beta (H-\mu
N)]^\frac{q}{1-q}=\{\texttt{Tr}[1-(1-q)\beta'
H_n]^\frac{q}{1-q}[1+(1-q)\beta'\mu']^\frac{q}{1-q}
\}^N=(\texttt{Tr}e^{-q\beta(h_n-\omega)})^N=z^N$ and where $z$ is one-particle
partition function. Just as for canonical ensemble, this exponential distribution can
be shown to be the result of the maximization of $S$ under the constraint
$\bar{N}=\texttt{Tr}\rho^qN$ in addition to Eq.(\ref{15}) and incomplete normalization.
Now Eq.(\ref{17b}) can be written as
\begin{eqnarray}                                        \label{19}
\bar{n}_k=\texttt{Tr}\rho^q n_k=\frac{1}{e^{q\beta'(\epsilon_k-\omega)}\pm1}
=\frac{1}{[1+(q-1)\beta' e_k]^{\frac{q}{q-1}} [1-(q-1)\beta'\mu']^{\frac{q}{q-1}}\pm1}
\end{eqnarray}
where $\epsilon_k$ is the eigenvalue of $h_n$. From Eq.(\ref{17b}), we see
that, for free particles ({\it in the sense that we do not write the energy of
interaction between particles in the hamiltonian and let it be ``absorbed" in
the nonextensive part of energy $H_c$ and related to $q$ different from
unity}), we have to set $q<1$ to ensure positive $[1+(q-1)\beta(H_n-\mu)]$ for
fermions when $T\rightarrow 0$. This means that, at low temperatures, there
will be few fermions beyond Fermi-energy. For bosons with $\mu<0$, we have to
put $q>1$.

It is straightforward to write $S=\ln Z+\beta' u+\beta'\omega\bar{N}$ and, with
the method employed in references \cite{Wang00,Wang01}, to show that, for a
system $C$ containing two equilibrium systems $A$ and $B$ satisfying
$\bar{N}(C)=\bar{N}(A)+\bar{N}(B)$, $\beta'(A)=\beta'(B)$ and
$\omega(A)=\omega(B)$ thus $\mu'(A)=\mu'(B)$, $\mu(A)=\mu(B)$ and
$\beta(A)=\beta(B)$ or $T(A)=T(B)$. The zeroth law is verified. One may ask why
we identify $\beta$, instead of $\beta'$, to real temperature. The possible
reasons are the following : 1) $\beta$ is the Lagrange multiplier of the
constraint on real energy in entropy maximization; 2) if $\beta'$ is inverse
temperature, then $\mu'$ must be chemical potential, which makes it impossible
to get distribution Eq.(\ref{18}) by real entropy maximization with the
constraint on $\bar{N}$; 3) $e_f$ would be different from the chemical
potential $\mu'$ and equal to $\frac{\mu'}{1+(1-q)\beta'\mu'}$ which inevitably
drops to zero when $T\rightarrow 0$.

Now we will focus the discussion on 2D fermion distribution. According to
Eq.(\ref{17b}) the Fermi energy $e_{f_0}$ at $T=0$ is given by
$e_{f_0}=\frac{2\pi\hbar^2\sigma}{m}$ where $\sigma$ is the particle density
and $\hbar$ Planck constant.

When $T>0$, the summation $\bar{N}=\sum_k\bar{n}_k$ can not be calculated for
arbitrary $q$ to give explicit expression of $e_f$. So we have recourse to
numerical calculation of $q$-dependence of $e_f$ for given temperatures (Figure
1) and $T$-dependence for given $q$ values (Figure 2). We see that, for the
approximate distributions functions (ADF) $n_k=1/\{[1+(q-1)\beta
(e_k-\mu)]^{\frac{1}{q-1}}+1 \}$ deduced from Eq.(\ref{17}) with factorization
approximation and additive energy\cite{Buyu93}, $e_f$ depends only slightly on
$q$. On the other hand, the $e_f$ of IS in the present work shows a strong
increase with decreasing $q$ up to two times $e_{f_0}$ of conventional
Fermi-Dirac distribution (CFD). This $e_f$ increase has been indeed noticed
through numerical calculations for strongly correlated heavy electrons on the
basis of tight-binding Kondo lattice model\cite{Corelec2,Corelec3}. In Figure
2, we show that $e_f$ of IS does not monotonically decrease with increasing
temperature, in contrast with the $e_f$ behavior of CFD. This kind of non
monotonic temperature evolution of $e_f$ was also seen through numerical work
for correlated electrons in two-dimensional $t-J$ model\cite{Puti98a}

The IS distribution given by Eq.(\ref{17b}) is plotted in Figure 3 for $T=100
K$. The particle density $\sigma$ is chosen to given $e_{f_0}=1$ eV. As
expected, the distribution changes drastically with $q$. When $q$ decreases, we
notice a flattening of the distribution with always a sharp drop of occupation
number $n$ at $e_f$ which increases. This flattening is also noticed in
numerical calculations for strong coupling electrons\cite{Corelec2,Corelec3}.

The grand canonical partition function $Z$ can be calculated to give the
$q$-deformed grand potential $\bar{\omega}=-T'\ln
Z=\frac{T'}{q}\sum_k\ln(1-\bar{n}_k)$ as usual, where $T'=1/\beta'=T+(1-q)\mu$.
The grand potential is given by :
\begin{eqnarray}                                        \label{21}
\Omega=\frac{e^{(q-1)\beta'\bar{\omega}}-1}{(q-1)\beta}
=T\frac{\prod_k(1-\bar{n}_k)^{\frac{q-1}{q}}-1}{(q-1)}.
\end{eqnarray}

In this $q$-deformed extensive formalism, Euler theorem applies just as in BGS.
So we have $g=\omega \bar{N}=u+pV-T'S$. Compare this with $S=\ln Z+\beta'
u-\beta'\omega \bar{N}$ to obtain the following equations of state :
\begin{eqnarray}                                        \label{22}
pV=T'\ln Z=\frac{T'}{q}\sum_k\ln(1-\bar{n}_k).
\end{eqnarray}

\section{Discussion and conclusion}
The formalism of NSM presented here is required by the existence of
thermodynamic equilibrium or by Eq.(\ref{m1}) for nonextensive systems
described by Tsallis entropy. Theoretically the formalism is self-consistent.
Experimental or numerical evidences are needed to verify the thermodynamic
relations. In this framework, all the successful applications of NSM conforming
with Eq.(\ref{m1}) are still valid. But the approximate applications carried
out for many-body systems using additive energy as exact hamiltonian (not
consistent with the existence of thermodynamic equilibrium) should be carefully
reviewed.

As mentioned above, we have noticed similar properties between the IS fermion
distribution and that of strongly correlated electrons\cite{Corelec2,Corelec3}.
This similarity shows the merit of NSM in describing strong interacting
systems. On the other hand, we noticed that a flattening of $n$ drop at $e_f$
with increasing correlation, observed experimentally and numerically with
weakly correlated electrons\cite{Corelec2,Corelec3,Corelec4,Ronn98,Puti98b}, is
absent within NSM fermion distributions which always show very sharp $n$ drop
at $e_f$ at low temperatures. A detailed study of this problem will be
presented in another paper of ours.

It is worth mentioning that, in the present work, energy is nonadditive to
satisfy the requirement of thermodynamic equilibrium. Nonadditive energy can
happen if interaction is no longer of short-range and not localized only
between the containing walls of subsystems. But in the literature, there are
rarely explicit expressions of nonadditive energy. One of explicit examples is
the long-range Ising model\cite{Sala01} where the internal energy $U(N,T)$ may
proportional to $N^c$ ($c$ is a constant), instead of $N$, the number of spins
in the system. This energy can be shown to satisfy Abe's pseudoadditivity for
energy\cite{Wang02} if we choose f(U) to be proportional, e.g., to $N$ (with
$\lambda_H=0$) or to $e^N-1$ (with $\lambda_H=1$).

Indeed, theoretically, nonadditive physical quantities (energy, volume ... ) is
not evident within the statistics with complete distributions because all
possible states (all points in phase space) are counted and summed here. But
from the viewpoint of incomplete statistics, nonadditivity may be interpreted
as a consequence of incomplete summation of state points in phase space due to
the incompleteness of our knowledge about the physical systems\cite{Wang00}.

Nevertheless, the fact that the correlation energy $H_c$ of NSM depends on
temperature, as shown in Eq.(\ref{2}) or Eq.(\ref{2c}), is not an easy aspect
to be understood. A possible interpretation is that these nonadditive equations
are required or prescribed by thermal equilibrium with Tsallis entropy and so
naturally change with temperature. This implies that the effect of correlations
may depend on temperature.

Summing up, within the framework of {\it incomplete statistics}, it is argued
that the nonextensive thermostatistics should be based on the factorization of
compound probability suggested, not by ``independence" of noninteracting
systems, but by {\it existence of thermodynamic equilibrium in interacting
systems having Tsallis entropy}. So this factorization must be viewed as a
fundamental hypothesis of NSM and rigorously satisfied by all exact
calculations relative to equilibrium systems. On this basis, we have elaborated
an additive formalism of NSM based on the maximization of an additive deformed
entropy subject to constraints on additive particle number and additive
$q$-deformed energy. The IS quantum distributions of this formalism are
compared with the distributions previously obtained by using complete
probability and additive energy in factorization approximation and also with
numerical results for strongly correlated electrons. It is shown that some
effects of strong correlations : the flattening of the fermion distribution and
the sharp cutoff of occupation number at $e_f$ which shows strong increase with
increasing interaction, can be observed in IS fermion distribution with
decreasing $q$ value.

\newpage

{\Large Figures:}

\begin{figure}[p]

\caption{$q$-dependence of Fermi energy $e_f$ of IS quantum distribution in
present work and of ADF given by the factorisation approximation with additive
energy. The fermion density $\sigma$ is chosen to give $e_f^0 = 1$ eV for CFD
distribution at $T=0$. IS $e_f$ shows strong increase with decreasing $q$ up to
two times $e_{f_0}$. But ADF $e_f$ depends only slightly on $q$. We also notice
that the $T$-dependence of IS $e_f$ is not monotonic as shown in Figure 2.}
\end{figure}

\begin{figure}[p]

\caption{$T$-dependence of Fermi energy $e_f$ of IS quantum distribution in
present work. The fermion density $\sigma$ is chosen to give $e_f^0 = 1$ eV for
CFD distribution at $T=0$. The $T$-dependence of $e_f$ is in general not
monotonic, in contrast with the classical decreasing behavior of $e_f$ with
increasing temperature. We notice that, at low temperature, $e_f$ shows an
increase with increasing $T$.}
\end{figure}

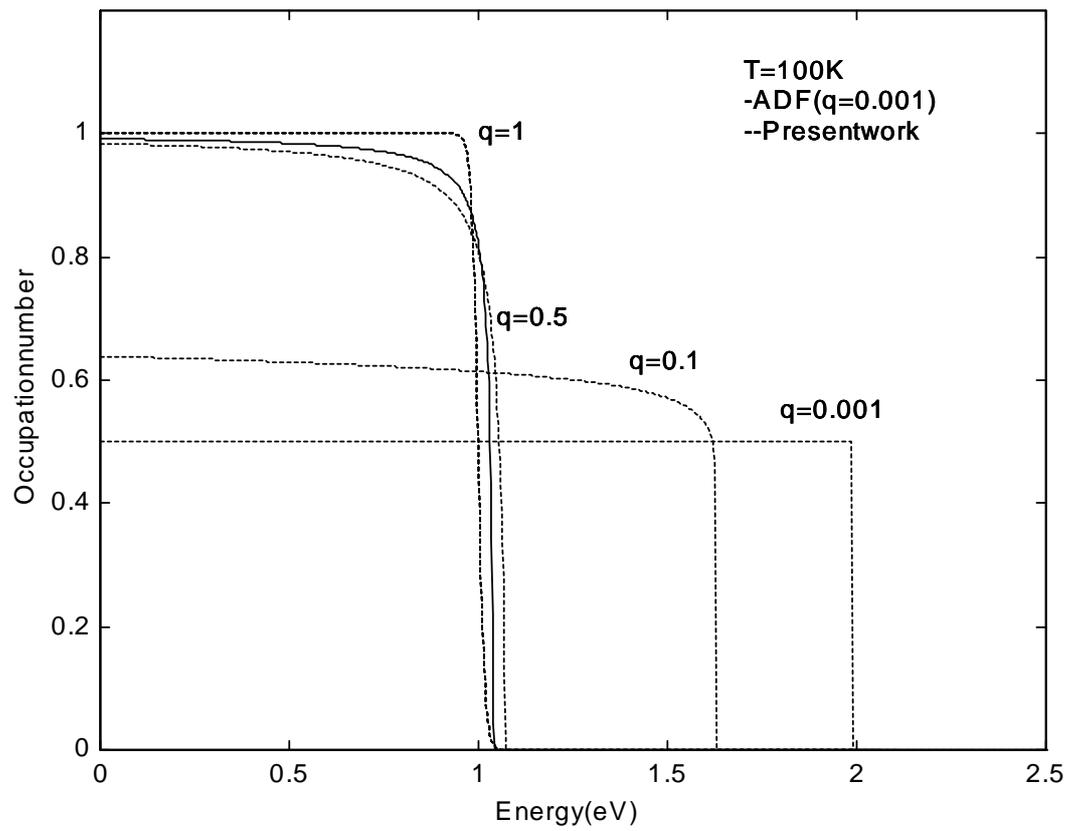
\begin{figure}[p]

\caption{Fermion distributions of ADF and of IS in present work. ADF
distribution is only slightly different from that at $q=1$ (CFD one) even with
$q$ very different from unity. The IS distribution of present work changes
drastically with decreasing $q$. As $q\rightarrow 0$, the occupation number
tends to 1/2 for all states below $e_f$ which increases up to 2 times
$e_{f_0}$.}
\end{figure}


\begin{thebibliography}{99}
\bibitem {Entropy}
Tsallis entropy $S_q$ can be given by averaging $I_g$ in several ways according
to the normalization of $\rho$ : 1) Supposing complete distribution with
$\texttt{Tr}\rho=1$, let $g=1-q$, we have $S_q=k\texttt{Tr}\rho I_g
=k\texttt{Tr}\rho\frac{(1/\rho)^{1-q}-1}{1-q}
=-k\texttt{Tr}\frac{\rho-\rho^q}{1-q}$\cite{Tsal01a}. $k$ is Boltzmann
constant. 2) Supposing incomplete distribution with $\texttt{Tr}\rho^q=1$, let
$g=q-1$, we have $S_q =k\texttt{Tr}\rho^q\frac{(1/\rho)^{q-1}-1}{q-1}
=-k\texttt{Tr}\frac{\rho-\rho^q}{1-q}$\cite{Wang00}. This calculation also
holds if we suppose complete distribution and the unnormalized average of an
operator $x$ : $\bar{x}=\texttt{Tr}\rho^q x$\cite{Tsal88}.
\end{thebibliography}
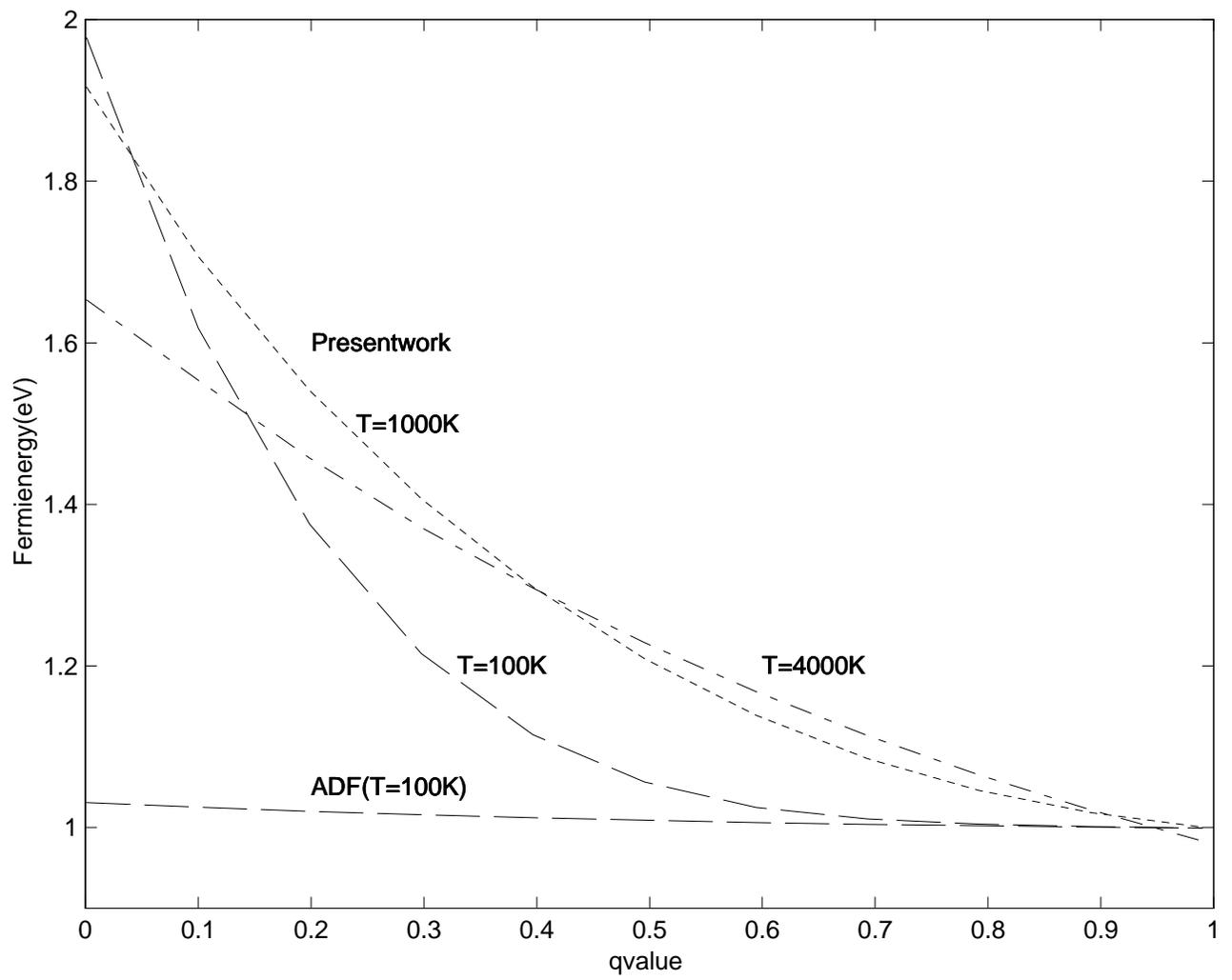
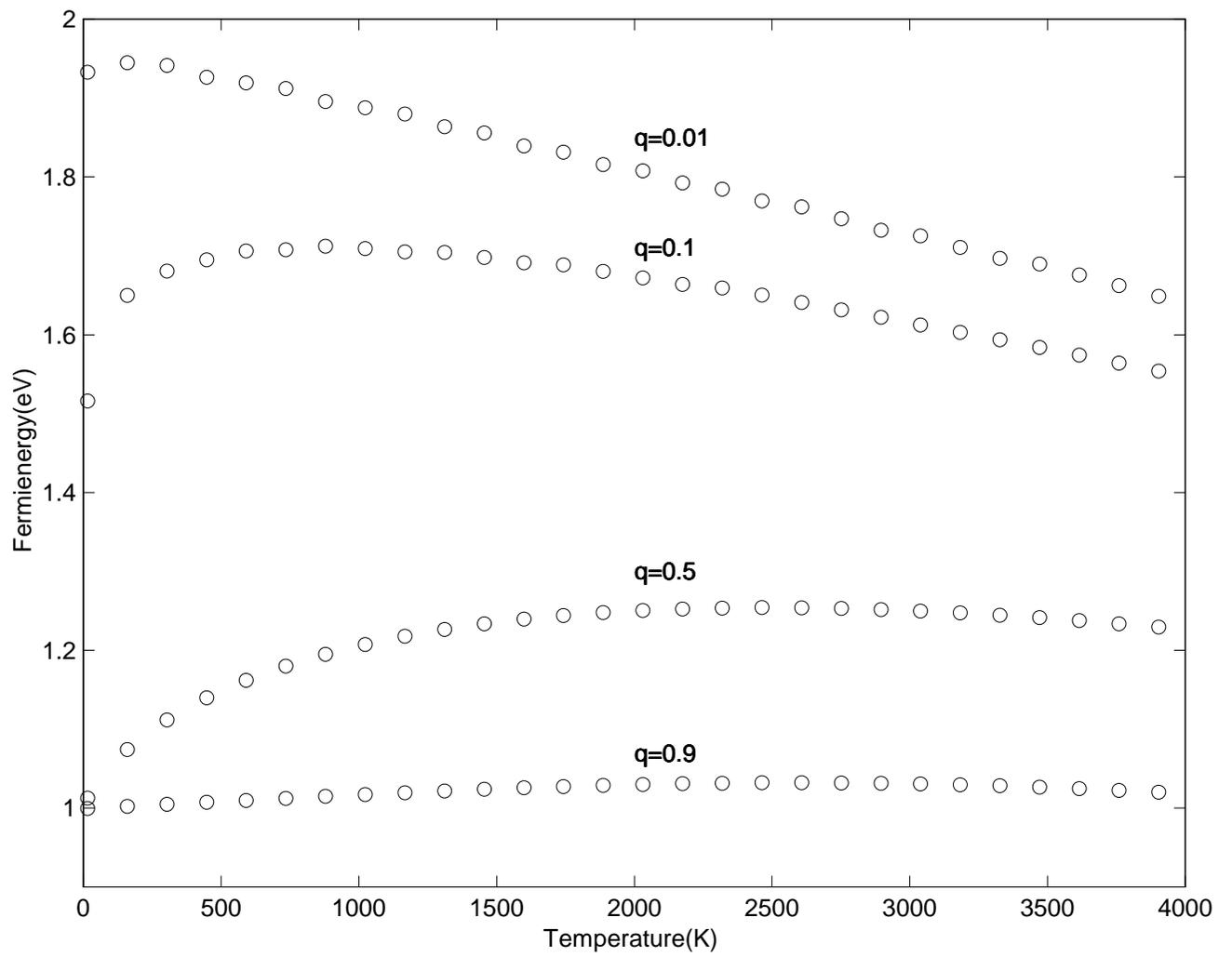
\end{document}